\documentclass[conference]{IEEEtran}
\IEEEoverridecommandlockouts
\usepackage{cite}
\usepackage{amsmath,amssymb,amsfonts}
\usepackage{algorithmic}
\usepackage{graphicx}
\usepackage{textcomp}
\usepackage{xcolor}
\usepackage{tabularx}
\usepackage{tabularx}
\usepackage{array}
\usepackage{graphicx}
\usepackage{adjustbox}
\usepackage{authblk} 

\def\BibTeX{{\rm B\kern-.05em{\sc i\kern-.025em b}\kern-.08em
    T\kern-.1667em\lower.7ex\hbox{E}\kern-.125emX}}

\begin{document}

\title{AuthAttLyzer-V2: Unveiling Code Authorship Attribution using Enhanced Ensemble Learning Models and Generating Benchmark Dataset\\
}

\author[1]{Bhaskar Joshi}
\author[2]{Sepideh HajiHosseinKhani}
\author[2]{Arash Habibi Lashkari}

\affil[1]{International Institute of Information Technology, Hyderabad, India }
\affil[2]{Behaviour-Centric Cybersecurity Center (BCCC), School of Information Technology, York University, Toronto}

\maketitle

\begin{abstract}
Source Code Authorship attribution (SCAA) is crucial for software classification because it provides insights into the origin and behavior of software. By accurately identifying the author or group behind a piece of code, experts can better understand the motivations and techniques of developers. In the cybersecurity era, this attribution helps trace the source of malicious software, identify patterns in the code that may indicate specific threat actors or groups, and ultimately enhance threat intelligence and mitigation strategies. This paper presents “AuthAttLyzer-V2,” a new source code feature extractor for SCAA, focusing on lexical, semantic, syntactic, and N-gram features. Our research explores author identification in C ++ by examining 24,000 source code samples from 3,000 authors. Our methodology integrates Random Forest, Gradient Boosting, and XGBoost models, enhanced with SHAP for interpretability. The study demonstrates how ensemble models can effectively discern individual coding styles, offering insights into the unique attributes of code authorship. This approach is pivotal in understanding and interpreting complex patterns in authorship attribution, especially for malware classification

\end{abstract}

\begin{IEEEkeywords}
Source Code Analysis, Source Code Authorship Attribution, Authorship Attribution, Ensemble Learning,
Explainable Learning
\end{IEEEkeywords}

\section{\textbf{Introduction}}
The source code surpasses its foundational logical structures and algorithms; it serves as a reflective canvas, capturing the unique creative essence of the programmer (Holcombe, 2019). Just as an author’s unique handwriting characterizes their work, one might ponder whether the individuality of a programmer can be identified within the strokes of code they write (Onan, 2018). This fascinating question lies at the heart of an in-depth investigation into the concept of coding styles serving as a digital signature, thereby laying the groundwork for the discipline of Author Attribution (Gama et al., 2018). 

In exploring the existing literature on authorship attribution, we have dedicated our efforts to systematically categorize the wealth of research into coherent approaches and pinpoint areas ripe for further inquiry. This endeavor illuminates the pivotal role of coding style analysis, as exemplified by studies like (Caliskan-Islam et al., 2015; Frantzeskou et al., 2006; Simko et al., 2018). These researchers have pioneered techniques for identifying authors through the distinct patterns in their coding, navigating through the complexities of dataset diversity and the susceptibility of their methods to adversarial interference. Additionally, language-specific research (Barlas and Stamatatos, 2020; Ramezani et al., 2013) has explored authorship attribution with a focus on particular programming languages, revealing both the potential and limitations of these approaches. Advances in machine learning (ML) and deep learning (DL) have further enriched authorship attribution techniques, offering sophisticated methods for identifying authors, even in the context of binary code and malware analysis (Mechti and Almansour, 2021; Krsul and Spafford, 1997; Rosenblum et al., 2011). Despite these advancements, the research also points out the need for improved defenses against adversarial attacks, suggesting an ongoing evolution in the field of authorship attribution (Quiring et al., 2019; Abhishek et al., 2022).

As highlighted in the preceding text, the field of authorship attribution has seen extensive research. Still, it continues to grapple with challenges such as scalability (Ding and Samadzadeh, 2004), coding styles (Caliskan-Islam et al., 2015), evaluation metrics (Amin et al., 2020), and generalizability (Barlas and Stamatatos, 2020). These issues, especially when related to tasks like malware authorship identification (Shin et al., 2020), underscore the critical importance of this domain. In response to these challenges, this study aims to advance the field by developing an enhanced version of AuthAttLyzer, incorporating expanded features. Moreover, we propose an interpretable algorithm designed to elucidate the decision-making process by integrating SHAP with various machine-learning techniques.

Additionally, this research introduces a novel dataset, BCCC-AuthAtt-2024, specifically curated for our experiments. This dataset serves as the foundation for testing a cutting-edge Authorship Attribution methodology, marking a significant step forward in refining and broadening the scope of authorship attribution studies. Employing this new dataset, our study encompassed a broad analysis across 24,000 instances, spanning eight author categories. The methodology involved the utilization of Random Forest, Gradient Boosting, and XGBoost algorithms, both with and without the integration of SHAP for interpretability, to identify the most effective model for authorship attribution. The results of these comprehensive experiments, highlighting metrics such as precision, recall, accuracy, and F1-score, illustrate the efficacy of our feature selection strategy. Moreover, they reveal the advantages of employing a hybrid approach to achieve enhanced performance in the analysis of Smart Contracts. 

Our contributions in this study are multi-faceted, providing significant advancements in the realm of authorship attribution in C++ source codes:
\begin{itemize}
    \item To propose a novel feature taxonomy, spotlighting the attributes derived from C++ source code, facilitating the development of a more nuanced and efficacious approach to authorship attribution.
    \item To introduce an enhanced iteration of AuthAttLyzer, building upon the foundation laid by AuthAttLyzer (version 1), by integrating additional pertinent features for a more thorough grasp of the subject at hand.
    \item To create a new C++ source code Dataset namely BCCC-VolSCs-2023, by having source codes from 8 different authors, with the primary objective of implementing the proposed model on a substantial sample size of 24,000 samples.
    \item To develop and implement an innovative authorship attribution model that incorporates SHAP for enhanced explainability, utilizing the features of AuthAttLyzer version 2 to accurately determine code authorship.
\end{itemize}

The structure of the remainder of this paper is as follows. Section 2 delves into the existing literature within this research field and identifies the limitations inherent in current solutions. Section 3 is devoted to an in-depth discussion of our proposed model. Section 4 outlines the extensive experiments, results, and implementation of our model. Section 5 thoroughly analyzes and discusses. Finally, Section 6 presents our conclusions and suggests directions for future work.

\section{\textbf{Literature Review}}
This section provides a summary of previous research works on authorship attribution and their proposed solutions. In doing so, we aim to categorize previous works into different branches. The final subsection identifies the gaps and shortcomings within the existing literature and outlines which of them we will cover in this research.

Exploring the first category, researchers such as (Caliskan-Islam et al., 2015; Frantzeskou et al., 2006; Simko et al., 2018) have focused on authorship attribution through the analysis of coding styles. Initially, (Caliskan-Islam et al., 2015) highlighted the distinctiveness of coding styles as unique identifiers for programmers, particularly emphasizing C/C++ languages. Their Code Stylometry Feature Set (CSFS) surpassed previous methods in accuracy for attributing authorship. However, issues were raised regarding the dataset’s representativeness and practical implementation’s feasibility.

Continuing in this category, (Frantzeskou et al., 2006) introduced the Source Code Author Profiles (SCAP) method, employing byte-level n-gram profiles that were effective and language-independent, even with sparse training data. SCAP proved resilient against the absence of comments in source code and could accurately identify authors in Java and C++ with brief code samples. Yet, there were concerns about its ability to capture coding styles’ semantic and logical elements across various languages. Recent progress has also strengthened classifiers for source code attribution. (Simko et al., 2018) investigated situations where adversaries manipulate code to cause misclassification, either by mimicking another author’s style to hide their identity or falsely implicating others. Their study revealed the susceptibility of these systems to manipulation by non-expert adversaries. To craft more robust classifiers, they researched with C/C++ programmers to evaluate adversary strategies and countermeasures by human analysts and suggested features to improve system resilience against adversarial techniques.

Moving to the next category, (Ramezani et al., 2013; Barlas and Stamatatos, 2020) shifted their focus to language-specific approaches for authorship attribution. In 2013, (Ramezani et al., 2013) explored Persian corpora. They assessed textual features, including lexical, character, syntactic, and semantic categories, emphasizing the reliability of features related to used words and verbs, particularly NLP-based ones. However, the study’s exclusive focus on Persian limited generalizability and a deeper analysis of feature interplay and real-world applicability could provide more insights. Moreover, the scenarios explored may not fully represent diverse authorship styles or unstructured texts, and a lack of detailed error analysis hinders complete understanding. Following this idea in 2020, (Barlas and Stamatatos, 2020) addressed cross-domain attribution challenges crucial for cybersecurity, digital humanities, and social media analytics. Experiments on various text genres showed promising results, highlighting its effectiveness in cross-domain scenarios. They also emphasized the significance of the normalization corpus in improving attribution accuracy.

Whereas previous research categories concentrated on the analysis of coding styles and language-specific attributes, a group of researchers introduced a pioneering concept in authorship attribution through syntactic and semantic analysis. In 2010, (Chen et al., 2010) introduced a semantic approach that identifies authorship by analyzing program data flows, which is especially useful for detecting code theft. Their findings demonstrated resilience even in the face of code alterations. A few years later, (Sidorov et al., 2014) introduced an innovative method that generates n-grams by tracing paths within syntactic trees. Their research, utilizing a corpus from Project Gutenberg, demonstrated that sn-grams could surpass traditional n-grams in certain conditions, with Support Vector Machines (SVM) achieving perfect accuracy in specific instances. However, this method introduces challenges such as dependency on the language and focusing on literary texts from a specific period, prompting questions about its effectiveness across various literary styles and genres.

With the introduction and increasing application of ML and DL, several significant approaches have emerged in utilizing these technologies for authorship attribution. (Mechti and Almansour, 2021) underscored the importance of pattern size in coding analysis, building upon earlier work by (Krsul and Spafford, 1997) and (Ding and Samadzadeh, 2004) in identifying features for authorship attribution in C and C++ code. Later, focusing on binary features, the challenge intensified in binary code analysis, where the impact of the compilation process on a programmer’s style became a complex issue. (Rosenblum et al., 2011) tackled this by framing it as a ML challenge, proposing methods to identify authors within program binaries. Their experiments showed that a programmer’s unique style could be detected within binary code, highlighting the importance of authorship attribution across various domains. (Fabien et al., 2020) introduced BertAA, an advanced DL-based method for Authorship Attribution, proving its efficacy on datasets such as Enron Email, Blog Authorship, and IMDb. They also demonstrated the benefits of incorporating stylometric and hybrid features into an ensemble model for enhanced performance.

Proceeding in this category, (Zafar et al., 2020) developed a deep metric learning approach for SCAA, achieving notable accuracy and scalability, as demonstrated on the Google Code Jam dataset. These developments emphasize the potential of DL in authorship attribution, especially in identifying programmers through stylistic nuances in code, which is vital for software theft detection, digital forensics, and malware analysis. (Lee and Cho, 2022) identified malware authors through automated analysis, employing six distinct machine-learning algorithms. Features selected for author identification included Runtime Modules and Kernel32.dll API, extracted from the automated analysis, resulting in improved accuracy and efficiency. Building on the advancements in DL and the insights from the previous category, (Amin et al., 2020) created a DL model specifically for detecting Android malware behaviors. This development highlighted the significant potential of DL to bolster malware detection capabilities.

Exploring the subsequent category, which focuses on binary source code for authorship attribution, several pioneering proposals have surfaced over time. In response to Advanced Persistent Threats (APTs) utilizing malware, (Shin et al., 2020) developed a genetic algorithm-based framework for analyzing malware. This innovative framework identifies unique features within binary and source code, achieving an 86\% accuracy rate in classifying authors. Further advancements in malware detection were made by (Benthin, 2022), who investigated malware binaries to improve authorship attribution methods.

The final category enriches the discussion on the robustness of authorship attribution techniques, with (Quiring et al., 2019) introducing a strategy to undermine SCAA by employing adversarial examples. This approach highlights the vulnerability of current attribution models to intentional attacks, emphasizing the urgent need for the development of more resilient defense mechanisms. (Abhishek et al., 2022) tackled the vulnerability of SCAA models to gradient-based attacks and universal perturbations. They proposed a novel architecture that employs defensive distillation to improve SCAA’s defenses against adversarial attacks, managing to achieve up to 95\% accuracy on unmodified code files.

\subsection{Synthesis of Previous Works}
A closer examination of previous works on authorship attribution highlights various areas where limitations become apparent:
\begin{itemize}
    \item \textbf{Dataset Dependency:} The substantial achievements demonstrated by (Caliskan-Islam et al., 2015) heavily rely on the Google Code Jam dataset, which raises questions regarding the universality and generalizability of the proposed models.
    \item \textbf{Scalability Concerns:} The literature reveals a gap in exploring the computational demands and scalability of authorship attribution methods, leaving their practical industrial application uncertain (Barlas and Stamatatos, 2020; Mechti and Almansour, 2021; Ding and Samadzadeh, 2004).
    \item \textbf{Language-Specific Approaches:} The dependence on language-specific features in some studies suggests limitations in their applicability to diverse programming environments, necessitating cross-language validation (Sidorov et al., 2014).
    \item \textbf{Evolving Coding Styles:} The dynamic nature of coding practices and the evolution of individual styles are not fully addressed, potentially undermining the long-term accuracy of attribution models (Caliskan-Islam et al., 2015; Frantzeskou et al., 2006; Simko et al., 2018).
    \item \textbf{Model Evaluation Metrics:} There is an observed emphasis on accuracy, with other critical evaluation metrics such as precision, recall, and F1 scores receiving less attention, which is vital for a rounded assessment of classification models (Amin et al., 2020; Quiring et al., 2019; Abhishek et al., 2022).
    \item \textbf{Code Analysis Limitations:} Combining binary and source code analysis for authorship attribution shows promise but lacks optimal accuracy, particularly in malware profiling. Despite innovative approaches like genetic algorithms, achieving satisfactory classification accuracy remains challenging. Further advancement of feature extraction is necessary to enhance the effectiveness of attribution methods in cybersecurity. (Benthin, 2022; Shin et al., 2020)
    \item \textbf{Linguistic Generalizability:} Concentrating on specific languages within certain studies could impede extending their results to other linguistic contexts, underscoring the need for studies encompassing a broader linguistic spectrum (Ramezani et al., 2013; Barlas and Stamatatos, 2020).
    \item \textbf{Code Manipulation Resilience:} While some methods, like SCAP, show resilience to certain forms of code manipulation, their robustness against a wider array of alterations has not been thoroughly investigated (Frantzeskou et al., 2006).
\end{itemize}

In this study, we address objectives 1, 2, 4, 5, 6, and 8 through the development of a newly enhanced feature extractor, AuthAttLyzer version 2, designed to improve authorship attribution by expanding its feature set. We also design and build an explainable AI model, training it on the newly compiled BCCC-AuthAtt-2024 dataset.

\section{\textbf{Proposed Model}}
This section introduces the proposed model architecture for author attribution in C++ code. Initially, we detail the step-by-step methodology employed in the design of our model architecture. Subsequently, we compare two leading interpretability techniques, Shapley Additive Explanations (SHAP) and Local Interpretable Model-agnostic Explanations (LIME), highlighting their advantages and relevance to our scenario. Following that, we offer an extensive overview of the features incorporated into our model, including layout, dynamic, style, N-gram, and graph-based elements crucial for identifying the distinctive coding styles of individual authors.

\subsection{Model Architecture}
This study employs a detailed methodology for attributing authorship within the realm of SCAA, organized into several crucial phases, each designed to capture and analyze the distinctive coding patterns of various authors:
\begin{itemize}
    \item \textbf{Collection of Individual Source Code Corpora:} The initial phase focuses on gathering individual source code datasets. Leveraging the open-source Codeforces platform, we compile various coding examples to encompass various programming styles.
    \item \textbf{Data Pre-Processing:} The raw code samples undergo a meticulous pre-processing routine following collection. This involves cleaning and structuring the data, setting the stage for effective feature extraction and further analysis.
    \item \textbf{Feature Extraction from Source Code (AuthAttLyzer-V2):} At the heart of our approach is extracting distinctive features from the source code. We pinpoint and extract 54 key features that reflect the unique coding styles of individuals, encompassing lexical, syntactic, and structural dimensions. These features form the foundation of our author profiling framework.
    \item \textbf{Ensemble Models for Author Attribution:} Our methodology utilizes ensemble ML techniques, specifically Random Forest and Gradient Boosting models, renowned for their efficacy in intricate classification scenarios. These models are adept at uncovering nuanced coding patterns distinct to each author. By integrating SHAP (SHapley Additive Explanations), we achieve high predictive accuracy and ensure transparency in how model decisions are made. Combining ensemble models with explainable AI methods is crucial for dissecting the complex features that define an author’s coding style.
    \item \textbf{Model Validation on Data:} The concluding phase tests the model’s predictive performance on a separate dataset of unseen code samples. This step is essential to assess the model’s generalizability and success in accurately attributing source code to the rightful authors.
\end{itemize}

Through this structured approach, our research aims to profile authors effectively based on their coding practices using advanced ML models.

\subsection{Model Selection}
In the pursuit of transparency and accountability within AI-powered decisions, Explainable AI (XAI) techniques such as SHAP (Shapley Additive Explanations) (Lundberg and Lee, 2017) and LIME (Local Interpretable Model-agnostic Explanations) (Mishra et al., 2017) play pivotal roles. These methodologies aim to illuminate the decision-making processes of complex models, providing case-specific insights essential for fostering trust and understanding in AI applications. By comparing and contrasting SHAP with LIME, we elucidate the reasons behind favoring SHAP for enhancing model interpretability, particularly in intricate tasks such as C++ code authorship attribution.

LIME offers interpretability by approximating the model locally using a simpler, interpretable model that captures the impact of significant features on specific decisions. Its principle is encapsulated in the formula:

\begin{equation}
E(x) = \underset{g \in G}{\mathrm{argmin}} \, L(f,g,\pi_x) + \Omega(g)
\end{equation}

Here, \(L\) represents the loss function measuring discrepancies between the surrogate model (\(g\)) and the original model (\(f\)), with \(\Omega(g)\) denoting the surrogate model’s complexity. LIME’s local surrogate models aim to balance fidelity to the original complex model with the simplicity of explanations.

SHAP, in contrast, builds on the theoretically grounded Shapley values from cooperative game theory, offering a more comprehensive and equitable approach to attributing feature importance. The SHAP formula is given as:

\begin{equation}
\phi_j(\text{val}) = \sum_{S \subseteq M \setminus \{j\}} \frac{|S|!(m - |S| - 1)!}{m!} [\text{val}(S \cup \{j\}) - \text{val}(S)]
\end{equation}

This formula calculates the marginal contribution of each feature across all possible feature combinations, ensuring a fair and transparent attribution of importance to each feature. SHAP’s adherence to properties such as efficiency, symmetry, dummy, and additivity further underscores its robustness in interpretability.

The preference for SHAP in our analysis, especially for the nuanced domain of C++ code authorship attribution, stems from its ability to provide global interpretability with local fidelity. SHAP not only aligns with the rigorous requirements of transparency but also excels in dissecting complex model predictions into understandable contributions from each feature, as seen in:

\begin{equation}
g(z') = \phi_0 + \sum_{i=1}^{m} \phi_i z'_i
\end{equation}

This explanation model, where \(z'\) represents the binary representation of feature presence, provides a detailed understanding of how each feature influences the model’s prediction. It showcases SHAP’s superior capability in handling the complexity and non-linearity inherent in C++ code structures, making it a preferred choice over LIME for enhancing model interpretability.

In essence, the strategic choice of SHAP over LIME for our task is justified by SHAP’s comprehensive framework that merges intuitive understanding with the equitable theoretical foundation of Shapley values. This not only ensures a profound understanding of the contribution of each feature to the model’s decision but also significantly advances the interpretability and trustworthiness of complex models in the field of AI and ML.

\subsection{Features Extraction}
Feature extraction is crucial for Authorship Attribution, serving as the base for identifying unique coding patterns of different authors. By analyzing code characteristics, we gain insights that improve author identification and contribute to Cybersecurity. This process is key to better understanding and managing source code, improving security, maintainability, and standards compliance. Below, we describe the extracted features using our new Author Attribution Analyzer tool, AuthAttLyzer-V2.

\subsubsection{Features Description}
\begin{figure}[h]
    \centering
    \includegraphics[width=1.1 \linewidth]{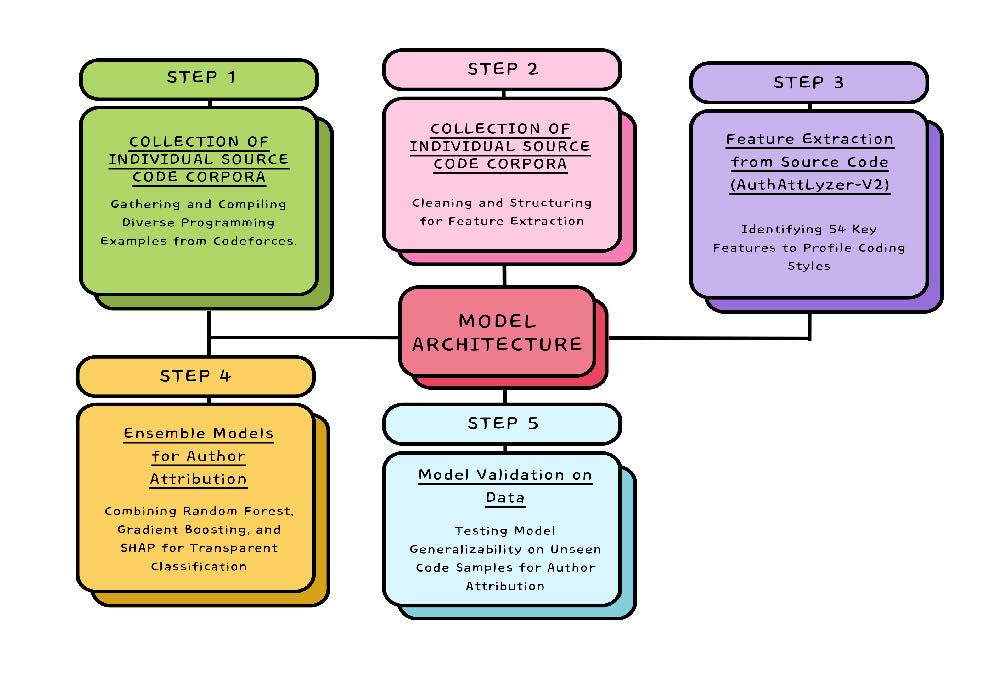}
    \caption{Taxonomy: Features Description}
    \label{fig:features}
\end{figure}

In this section, we delve into the detailed analysis of features designed to improve outcomes in Authorship Attribution, with a thorough evaluation set for the ”Experiments” section. Figure~\ref{fig:features} depicts the taxonomy of feature descriptions. These features are crucial for analyzing and understanding code, providing a range of practical applications:

\begin{itemize}
    \item \textbf{Code Cloning Detection:} Using character-level n-grams, we can accurately identify duplicate code fragments. When two segments of code exhibit a high similarity in their character-level n-grams, it suggests they are clones or direct copies of each other.
    \item \textbf{Code Style Analysis:} Word-level n-grams serve as a tool for evaluating conformity to coding standards and stylistic guidelines. Analyzing prevalent word-level patterns and their variations helps in pinpointing deviations from established coding norms.
    \item \textbf{Code Complexity Assessment:} The application of word-level n-grams aids in assessing the complexity of code. Identifying a frequent occurrence of complex constructs, such as nested loops or deep conditional statements, indicates a higher level of code complexity.
    \item \textbf{Code Anomaly Detection:} Both character and word-level n-grams are instrumental in spotting anomalies or potential security flaws. The detection of unusual character combinations or specific sequences of words (for instance, “eval” followed by a string) can prompt security warnings.
    \item \textbf{Language Identification:} Character-level n-grams are effective in determining the programming language of a piece of code. Each programming language has a unique n-gram profile, enabling precise classification of code snippets.
    \item \textbf{Code Summarization:} Word-level n-grams facilitate the creation of code summaries or documentation. Recognizing frequent constructs and comments allows for generating abstract descriptions outlining code functionalities.
\end{itemize}

The analysis of character and word-level n-grams yields significant insights into the structure, style, and content of source code. These features are indispensable across various domains, including code analysis, quality assessment, security examination, and automated documentation. They empower developers and code reviewers to delve deeper into codebases, identify potential issues, and make well-informed code refinement and enhancement decisions. The proposed feature extraction named AuthAttLyzer-V2 is an upgrade of our Authorship Attribution tool, enhanced better to identify coding styles and patterns in C++ programs. This version represents a significant improvement in helping to ensure software quality and security.

\begin{table}[htbp]
\renewcommand{\arraystretch}{2}
    \centering
    \resizebox{0.5\textwidth}{!}{
    \begin{tabularx}{0.8\textwidth} { 
       >{\large\centering\arraybackslash}p{0.07\textwidth} 
      | >{\large\centering\arraybackslash}p{0.22\textwidth} 
      | >{\large\centering\arraybackslash}p{0.4\textwidth}  }
    \hline
    \textbf{Layout Feature ID} & \textbf{Feature Focus} & \textbf{Feature Description} \\
    \hline
    tl\_1 & Operator Format & Measures the average number of whitespaces surrounding operators within the code. \\
    \hline
    tl\_2 & Frequency of left braces on the end of lines including 'if(condition)' & Examines how often left curly braces appear at the end of lines containing 'if(condition)' statements. \\
    \hline
    tl\_3 & Percentage of blank lines & Calculates the percentage of blank lines in the code, indicating code organization and readability. \\
    \hline
    tl\_4 & Average number of leading whitespaces per line & Quantifies the average indentation level across the code, reflecting code structure and formatting consistency. \\
    \hline
    tl\_5 & Format of for loop & Percentage of whitespaces between left and right parentheses of 'for(...)' to all 'for' loops.\\
    \hline
    tl\_6 & Format of for loop & Whether there is at least a whitespace between left and right parentheses of 'for(...)' \\
    \hline
    tl\_7 & ln(numTabs/length) & Log of the number of tab characters divided by file length in characters \\
    \hline
    tl\_8 & ln(numSpaces/length) & Log of the number of space characters divided by file length in characters \\
    \hline
    tl\_9 & ln(numEmptyLines/length) & Log of the number of empty lines divided by file length in characters, excluding leading and trailing lines\\
    \hline
    tl\_{10} & WhiteSpaceRatio & The ratio between the number of whitespace characters (spaces, tabs, and newlines) and non-whitespace characters \\
    \hline
    tl\_{11} & NewLineBeforeOpenBrace & Determines whether most code-block braces begin with a newline character. \\
    \hline
    tl\_{12} & TabsLeadLines & Assesses whether spaces or tabs are predominantly used for indentation in the code. \\
    \hline
  \end{tabularx}}
  \centering{
    \caption{Layout Features}
    \label{tab:my_label}}
\end{table}

    
    

\begin{table}[htbp]
\renewcommand{\arraystretch}{2}
  \centering
  
  \resizebox{0.5\textwidth}{!}{
    \begin{tabularx}{0.8\textwidth} { 
       >{\large\centering\arraybackslash}p{0.07\textwidth} 
      | >{\large\centering\arraybackslash}p{0.22\textwidth} 
      | >{\large\centering\arraybackslash}p{0.4\textwidth}  }
    \hline
    \textbf{Dynamic Feature ID} & \textbf{Feature Focus} & \textbf{Description} \\
    \hline
    $td_{1}$ & Extraction of Function Calls & Focuses on the extraction of function calls within the provided C++ code. \\
    \hline
    $td_{2}$ & Measurement of Total Memory Usage & Involves measuring the total memory usage of the C++ program  \\
    
    \hline
    $td_{3}$ & Measurement of Length of Disassembled Code & This feature measures the length of disassembled code generated by a compiler or disassembler tool. It compiles the provided C++ code to assembly language, reads the generated code, and reports its length, which can be indicative of the complexity of the code. \\
    \hline
    $td_{4}$ & Extraction of Time Information & Extracts time of execution information related to the execution of the C++ program. \\
    \hline
  \end{tabularx}}
  \caption{Dynamic Feature Descriptions}
  \label{tab:dynamic-feature-descriptions}
\end{table}

\begin{table}[htbp]
\renewcommand{\arraystretch}{2}
    \resizebox{0.5\textwidth}{!}{
    \begin{tabularx}{0.8\textwidth} { 
       >{\large\centering\arraybackslash}p{0.07\textwidth} 
      | >{\large\centering\arraybackslash}p{0.22\textwidth} 
      | >{\large\centering\arraybackslash}p{0.4\textwidth}  }
    \hline
    \textbf{Style Feature ID} & \textbf{Feature Focus} & \textbf{Feature Description} \\
    \hline
    $ts_1$ & Comment Lines & Measures the number of lines in the code that are comments, providing insights into code documentation. \\
    \hline
    $ts_2$ & Percentage of Comments & Calculates the percentage of lines that are comments relative to the total number of lines, indicating the level of code documentation. \\
    \hline
    $ts_3$ & Variable Name Length & Evaluates the average length of variable names in the code, reflecting naming conventions and readability. \\
    \hline
    $ts_4$ & Frequency of Loops & Comparing the frequency of the keyword “for” with that of “while”\\
    \hline
    $ts_5$ & Frequency of Control Statements & Compares the usage of "if" with that of "switch" statements, providing insights into conditional logic within the code. \\
    \hline
    $ts_6$ & Percentage of Static Global Variables & Calculates the percentage of lines containing static global variable declarations, indicating code structure and organization. \\
    \hline
    $ts_7-ts_9$ & Access Modifiers & Measures the use of access modifiers (public, private, protected), offering insights into code visibility and encapsulation. \\
    \hline
    $ts_{10}$ & Kinds of Operators & Identifies and lists the unique operators used in the code, revealing the diversity of operations performed. \\
    \hline
    $ts_{11}$ & Percentage of Operators & Calculates the percentage of lines containing operators relative to the total number of non-comment lines, indicating the complexity of operations. \\
    \hline
    $ts_{12}-ts_{15}$ & Percentage of Methods & Breaks down the percentage of lines with methods by their return types (int, char, void, String), providing insights into method distribution. \\
    \hline
    $ts_{16}$ & Percentage of Methods (Overall) & Measures the overall percentage of lines with methods relative to the total number of non-comment lines, revealing code structure. \\
    \hline
    $ts_{17}-ts_{20}$ & Variable Name Characteristics & Analyzes variable names for the presence of underscores, numbers, and uppercase starting variables, Camel-Case style \\
    \hline
    
    $ts_{21}$ & Number of Variables & Quantifies the total number of variables declared in the code, offering insights into code complexity and data management. \\
    \hline
    $ts_{22}$ & Percentage of Variables & Calculates the percentage of lines containing variable declarations relative to the total number of non-comment lines, revealing variable density. \\
    \hline
    $ts_{23}$ & Use of "goto" Statement & Indicates whether the "goto" statement is used in the code, reflecting code control flow. \\
    \hline
    $ts_{24}$ & Use of "auto" Statement & Indicates whether the "auto" statement is used in the code. \\
    \hline
    $ts_{25}$ & Use of "struct"  & Analyzes the use of "struct" \\
    \hline
  \end{tabularx}}
    \caption{Style Features}
    \label{Style_table}
\end{table}

\begin{table}[htbp]
\renewcommand{\arraystretch}{2}
  \centering

  \begin{adjustbox}{max width=0.5\textwidth}
    \begin{tabularx}{1.0\textwidth} { 
       >{\large\centering\arraybackslash}p{0.07\textwidth} 
      | >{\large\centering\arraybackslash}p{0.22\textwidth} 
      | >{\large\centering\arraybackslash}p{0.4\textwidth}  }
    \hline
  
    \textbf{Dynamic Feature ID} & \textbf{Feature Focus} & \textbf{Description} \\
    \hline
    $tn_1-tn_5$ & Character n-gram & Frequency of character-level n-grams\\
    \hline
    $tn_6-tn_9$ & Word n-gram & Frequency of word-level n-grams \\
    \hline
    
  \end{tabularx}
  \end{adjustbox}
  \caption{N-Gram based Features}
  \label{tab:NGram_Features}
  
\end{table}


  
    

\begin{table}[htbp]
\renewcommand{\arraystretch}{2}
  \centering

  \begin{adjustbox}{max width=0.5\textwidth}
    \begin{tabularx}{1.0\textwidth} { 
       >{\large\centering\arraybackslash}p{0.07\textwidth} 
      | >{\large\centering\arraybackslash}p{0.22\textwidth} 
      | >{\large\centering\arraybackslash}p{0.4\textwidth}  }
    \hline
    \textbf{Graph Based Feature ID} & \textbf{Feature Focus} & \textbf{Description} \\
    \hline
    $tg_1$ & Ast node bigram & Term frequency AST node bigrams \\
    \hline
    $tg_2$ & Ast node types & Term frequency of 58 possible AST node type excluding leaves\\
    \hline
    $tg_3$ & Ast node maxdepth & Maximum depth of an AST node \\
    \hline
    $tg_4$ & Code in Ast leaves & Term frequency of code unigrams in AST leaves \\
    \hline
    
  \end{tabularx}
  \end{adjustbox}
  \caption{Graph based Features}
  \label{tab:GrapbBased_Features}
  
\end{table}

\section{\textbf{Experiments}}
This section presents a concise yet comprehensive overview of our empirical investigation into C++ code authorship attribution. We commence with Strategic Algorithm Selection, elucidating the rationale behind our choice of ML models. This is followed by Model Performance Analysis, where we assess the efficacy of these models against a large dataset. Data Curation and Preparation detail the efforts to refine our dataset, while Metrics for Success define our evaluation criteria. Experimentation Framework outlines the setup for our research, leading to Analysis of Results, which reveals the key findings of our study. Each step is meticulously designed to ensure a robust and insightful exploration into identifying unique authorial styles in code.

\subsection{Data Preparation}
Our evaluation focuses on analyzing the source code from actual authors who have tackled numerous problems on the Codeforces platform, a hub for algorithmic challenges where participants can choose their preferred programming language for solution submission (Mirzayanov et al., 2020). This platform offers publicly accessible source codes from consistent contributors who are experienced in problem-solving over time. By selecting authors who have a history of genuine contributions and have tackled a diverse range of problems, we ensure the integrity of our dataset and include individuals who have demonstrated long-term engagement with the platform. This methodology allows us to capture any potential evolution in coding styles.

In terms of programming languages, C++ emerges as the most favored, followed by Java and Python. Our research into source code stylometry primarily targets C++ and C, chosen for their prevalence in competitive programming and the availability of efficient parsers for these languages. This study assesses the efficacy of multiple ML models in authorship attribution for C++ code using a newly generated dataset (BCCC-AuthAtt-2024) including 24,000 code samples (8 samples per author) from 3,000 authors. Our focus is on extracting a comprehensive array of features, including lexical, semantic, syntactic, and N-gram components, essential for identifying the unique coding styles of individual authors. The goal is to accurately predict an author’s identity based on their unique coding style.

\subsection{Evaluation Metrics}
The evaluation metrics for assessing the performance of designed and implemented explainable ML model, include several key statistical measures. These measures are based on outcomes categorized as True Positives (correct identifications of an author’s code), True Negatives (correct rejections of non-author code), False Positives (incorrectly attributing code to the specific author), and False Negatives (missing an author’s code). The primary metrics derived from these outcomes include:
\begin{itemize}
    \item \textbf{Accuracy:} This measures the overall correctness of the model, calculated as the ratio of correctly identified instances (both positive and negative) to all instances.
    \item \textbf{Precision:} This assesses the accuracy of the model’s positive predictions, indicating the proportion of positive identifications that are correct.
    \item \textbf{Recall:} This metric evaluates the model’s ability to identify all relevant instances, calculated as the ratio of correctly identified positive instances to all actual positive instances.
    \item \textbf{F1-score:} This is a harmonic mean of precision and recall, providing a single metric that balances both aspects to evaluate the model’s performance comprehensively.
\end{itemize}

The aforementioned metrics serve as the primary evaluation criteria for our model’s performance on the BCCC-AuthAtt-2024 dataset, ensuring a comprehensive understanding of its effectiveness in source code-based authorship attribution.

\subsection{Experimentation Setup}
The experiments for the implemented model were conducted using the Python programming language, version 3.8, on a system running Microsoft Windows 10 Home Version 20H2. Our primary development and evaluation environment included a workstation equipped with 16 GB of RAM, which provided a robust platform for handling the computational demands of the ML models. We employed several other libraries to aid in data preprocessing, model evaluation, and statistical analysis. NumPy (version 1.19) and Pandas (version 1.2) were used for data manipulation and analysis. Scikit-learn (version 0.24) was employed for its robust ML algorithms and tools, particularly for pre-processing steps and evaluation metrics.

\subsection{Model Performance}
Our research investigates the effectiveness of various ML models for attributing authorship using the BCCC-AuthAtt-2024 dataset. We meticulously evaluated six different models, each chosen for their relevance and potential contributions to ML’s theoretical and practical realms, aiming to determine their efficacy in code authorship attribution.
\begin{itemize}
    \item \textbf{Random Forest (RF):} RF stands as an ensemble learning method, valued for its resistance to overfitting in scenarios with high-dimensional data. Its capability to offer insights through feature importance analysis renders it an effective baseline model. RF was selected for its combination of simplicity, dependability, and interpretability—key attributes for preliminary model evaluation.
    \item \textbf{Random Forest with SHAP (RF-SHAP):} RF-SHAP enhances RF by incorporating SHAP explanations, shedding light on how individual features contribute to predictions. This model is favored for maintaining RF’s core advantages while significantly boosting interpretability, offering deeper insights into the coding styles specific to authors.
    \item \textbf{Gradient Boosting (GB):} GB stands out for its methodical approach of building models incrementally, enhancing precision in handling complex tasks and large datasets. It adeptly identifies non-linear relationships and complex interactions among features. GB was chosen for its potential to surpass RF in terms of accuracy.
    \item \textbf{Gradient Boosting with SHAP (GB-SHAP):} GB-SHAP marries the accuracy improvements of GB with SHAP’s explanatory depth, delivering accurate predictions and clarifying the impact of each feature. This model appeals for its blend of high precision and improved interpretability over the standard GB.
    \item \textbf{XGBoost:} XGBoost is recognized for its optimized gradient-boosting framework, which offers enhanced training speed and the possibility of achieving greater accuracy. It employs advanced regularization to minimize overfitting, making it particularly effective for large, complex datasets. XGBoost was selected for its robust performance in analyzing intricate patterns.
    \item \textbf{XGBoost with SHAP (XGB-SHAP):} XGB-SHAP combines XGBoost’s high accuracy with SHAP’s analytical depth, setting a new standard for precision and interpretability in our study. This model is ideally suited for research requiring both detailed model decision insights and superior accuracy, representing the epitome of performance and interpretability within our array of models.
\end{itemize}

Through this analysis, our goal is to discern the distinct coding styles of authors, leveraging advanced ML models to accurately attribute authorship, while also providing comprehensive insights into the decision-making processes of these models. Table~\ref{tab:performance_metrics} presents the performance metrics of the evaluated ML models using the BCCC-AuthAtt-2024 dataset.





\begin{table}[h]
\renewcommand{\arraystretch}{2}
\large
    \centering
    \caption{Performance metrics of ML models for code authorship attribution.}
    \resizebox{\columnwidth}{!}{
    \begin{tabular}{lcccc}
        \hline
        \textbf{Model} & \textbf{Accuracy (\%)} & \textbf{Precision (\%)} & \textbf{Recall (\%)} & \textbf{F1-score (\%)} \\
        \hline
        Random Forest & 67.5 & 68.2 & 66.7 & 67.4 \\
        Random Forest with SHAP & 78.4 & 79.1 & 77.7 & 78.4 \\
        Gradient Boosting & 69.7 & 70.4 & 69.0 & 69.7 \\
        Gradient Boosting with SHAP & 75.1 & 75.8 & 84.4 & 79.9 \\
        XGBoost & 70.8 & 71.5 & 70.1 & 70.8 \\
        XGBoost with SHAP & 81.2 & 81.9 & 80.5 & 81.2 \\
        \hline
    \end{tabular}
    }
    \label{tab:performance_metrics}
\end{table}


\section{\textbf{Analysis and Discussion}}
The study aimed to advance the field of authorship attribution by enhancing the AuthAttLyzer tool to its second version, AuthAttLyzer V2, which now incorporates 54 distinct features for the analysis of C++ code samples. To improve the model’s explainability, we integrated SHAP (SHapley Additive exPlanations) with various ML algorithms, providing insight into how each feature influences the model’s predictions. This methodology was applied to a newly established benchmark consisting of 24,000 C++ code samples from eight different authors, aiming to evaluate the model’s performance across multiple metrics rigorously.

\begin{figure}
    \centering
    \includegraphics[width=1 \linewidth]{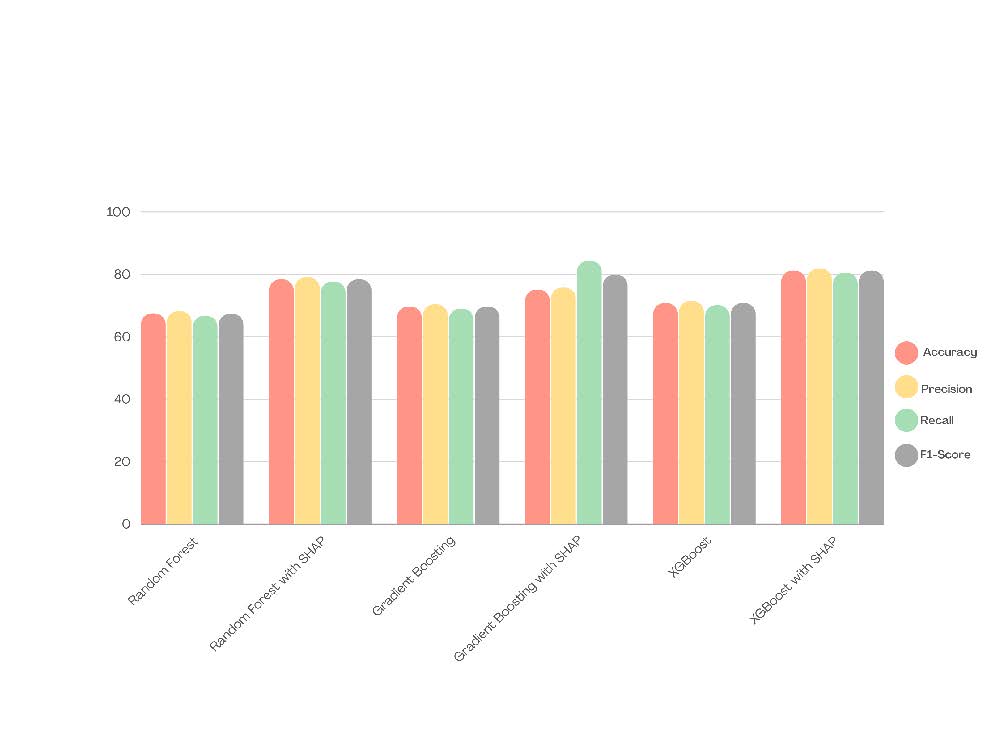}
    \caption{Comparative Performance Analysis of Different Models}
    \label{fig:performance_comparison}
\end{figure}

Our findings reveal that XGBoost with SHAP achieves a commendable balance in accuracy, precision, recall, and F1-score, as illustrated by a comprehensive comparison in Figure~\ref{fig:performance_comparison}. This equilibrium in performance metrics underscores the model’s robustness in attributing authorship accurately across a varied dataset, demonstrating its effectiveness in distinguishing among authors’ coding styles. The integration of SHAP bolstered the model’s predictive accuracy and its interpretability, enabling users to understand the rationale behind each attribution, a critical aspect for applications in both academic and security contexts.

This groundbreaking work with AuthAttLyzer V2 and its integration with an XGBoost model, enhanced by SHAP, sets a new benchmark in the field of authorship attribution. Beyond laying a solid foundation for future research to refine authorship identification methods, it significantly extends the applications of these technologies. Importantly, it introduces a powerful cybersecurity and malware analysis tool, where identifying malicious code authors can be crucial in tracking cyber threats and understanding attack methodologies. This work opens new avenues for research in both code authorship understanding and the broader context of cybersecurity, suggesting exciting directions for future investigations in code analysis.

\section{\textbf{Conclusion and Future Work}}
This study unveils AuthAttLyzer-V2, a cutting-edge feature extraction tool for SCAA, marking a significant advancement in accurately identifying code authors, particularly within the cybersecurity realm. AuthAttLyzer-V2 adopts a holistic approach, utilizing a meticulously selected combination of lexical, semantic, syntactic, and N-gram features to achieve this goal. Our analysis of 24,000 C++ source code samples from 3,000 authors demonstrates the power of combining Random Forest, Gradient Boosting, and XGBoost models with SHAP for enhanced interpretability. This approach effectively navigates the intricate patterns in C++ code, surpassing simpler algorithms and establishing a robust basis for identifying distinct coding styles.

The strategic use of XGBoost paired with SHAP results in an outstanding performance, demonstrated by accuracy, precision, recall, and F1-score rates all around 81.2\%, 81.9\%, 80.5\%, and 81.2\%, respectively. SHAP’s transformative impact in clarifying the factors driving model decisions enriches our understanding, corroborates our results, and offers valuable perspectives for ongoing and future applications. AuthAttLyzer-V2’s ability to accurately trace code to its authors sheds light on program origins and behaviors. This research advances the precision and practicality of authorship attribution and strengthens cybersecurity defenses, providing a robust basis for future explorations in this critical field.

Looking ahead, code authorship attribution offers exciting opportunities for growth and new ideas. By using the models we created in this study on more types of programming languages, we’re getting close to being able to profile any code universally. This could help reduce the differences between various coding languages. Further exploration into the depths of model features and their interactions through advanced interpretability techniques holds the potential to unravel the intricacies behind author identification. This could lead to a more nuanced understanding of authorship patterns and motivations. Applying interpretability tools like SHAP in real-world settings could revolutionize how developers and forensic investigators approach code authorship analysis, providing them unparalleled precision and confidence. This research has set the groundwork for understanding code authorship better, leading to a future where we can more effectively analyze and protect the safety and reliability of software worldwide.

\section*{\textbf{Availability}}
The source code for Authorship Attribution Analyzer “AuthAttLyzer-V2” and  the generated dataset “BCCC-AuthAtt-2024” will be made available on our website \cite{BCCC2024b}.

\section*{\textbf{Acknowledgements}}
The authors acknowledge the grant from Canada Research Chair - Tier II (\#CRC-2021-00340) and the Mitacs Global Research Internship (GRI) grant for supporting this research by providing the research internship opportunity.


\end{document}